\documentclass[12pt,a4paper]{article}
\usepackage{amsfonts,amsmath,amssymb,fullpage,graphicx,mathrsfs}
\usepackage{color,dashbox}

\newtheorem{corollary}{Corollary}[section]
\newtheorem{definition}[corollary]{Definition}
\newtheorem{example}[corollary]{Example}

\newtheorem{operation}[corollary]{Operation}
\newtheorem{proposition}[corollary]{Proposition}

\newtheorem{theorem}[corollary]{Theorem}

\newcommand{\remove}[1]{}
\newcommand{\qed}{\rule{.07in}{.1in}}
\newenvironment{proof}{\vspace{1ex}\noindent\textbf{Proof}\hspace{0.5em}}{\hfill\qed\vspace{1ex}}

\title{Dynamic Exponent Market Maker: Personalized Portfolio Manager and One Pool to Trade Them All}
\author{Wittawat Kositwattanarerk
\thanks{W. Kositwattanarerk is with the Department of Mathematics, Faculty of Science, Mahidol University, Bangkok 10400, Thailand and the Centre of Excellence in Mathematics, the Commission on Higher Education, Bangkok 10400, Thailand (e-mail: wittawat.kos@mahidol.edu).}}
\date{\today}

\begin{document}

\maketitle

\begin{abstract}
Decentralized exchange platforms such as Uniswap and Balancer operate on several pools where each pool contains two or more cryptocurrencies and constitutes direct trading pairs. The drawbacks here are that liquidity providing requires contribution of tokens in a specific proportion, and trading may require hopping between pools, hence increasing transaction fee and gas fee. We propose an automated market maker (AMM) protocol where liquidity providers can deposit any amount of tokens into the pool. The protocol will preserve the proportion of tokens by total value at the time of deposit and can be seen as a personalized self-balancing portfolio manager. In addition, since the invariant function is dynamic, all exchange pairs are executed from a single composite pool. Nevertheless, the scheme is vulnerable to flash loan attacks and must be used in conjunction with preventive measures.
\end{abstract}


\section{Introduction}

Blockchain has promising applications in many areas such as Internet of Things (IoT), supply chain, art and entertainment, healthcare, and obviously finance \cite{KHE,MMTL,PSS,USGB,ZMZZL}. While Bitcoin and its underlying infrastructure can be considered the world's first decentralized payment system, the term decentralized finance (DeFi) is nowadays used to refer to the post-Ethereum era where more advanced financial protocols such as lending and borrowing are made possible via smart contracts. Particularly, one of the cornerstones in a DeFi ecosystem is decentralized exchange (DEX), which allows a large number of users to trade cryptocurrencies without the need for a middleman.

Traditionally, large-scale exchanges such as a stock market require a centralized exchange (CEX). Here, buyers and sellers submit their orders which consist of asset prices and volumes to an intermediary who is in charge of keeping and maintaining an order book. A trade is settled if the there are matching orders. This approach contrasts sharply with blockchain ideologies since the process is centralized and participants of the trade have to entrust their funds to a third party. The disastrous collapses of Mt. Gox and FTX Trading Ltd., to name a few, serve as a bitter reminder of the risks associated with CEX.

On the other hand, decentralized exchange (DEX) is one of the remarkable innovations in DeFi. Instead of relying on the conventional market order book or a central authority, DEX is a set of smart contract algorithms where a trade is made against a pool of two or more cryptocurrencies. Typically, liquidity providers deposit two different types of tokens into a pool. At any time, a user can make a trade by swapping one type of token for another where the exchange rate is determined by a mathematical formula known as automated market maker (AMM). Once a trade is executed, liquidity providers earn a small transaction fee. Since there is an opportunity for a financial gain, providing funds for DEX becomes a new type of investment unique to DeFi called ``liquidity mining'' or ``yield farming''. Here, we refer interested readers to \cite{XPCF} for a comprehensive survey on decentralized exchange, to \cite{LWCLL} for a discussion on hybrid exchange, and to \cite{XF} for an overview on yield farming.

The aforementioned concepts of DEX come into existence with the launch of Uniswap \cite{A,AZR,AZSKR}, the first Ethereum-based and currently largest \cite{D} decentralized exchange. It utilizes a simple yet elegant formula known as the constant product market maker (CPMM) $xy=k$. While many exchanges follow suit and adopt the same formula \cite{P,Q,S}, many alternatives exist and have since served different purposes. For example, StableSwap (Curve) uses adaptive formula and is particularly suitable for exchanging stablecoins \cite{E}. Balancer uses weighted exponent so that the pool doubles as a self-balancing portfolio manager \cite{MM}. Nowadays, decentralized exchanges have become a multi-billion dollars industry \cite{C}, and the underlying mechanism has been extensively studied \cite{XPCF,AEC,AKCNC,TPL}. Particularly, the use of liquidity pool as an investment is studied in \cite{F,WK}, and the constant product formula is generalized in \cite{FK,KFG}.

Decentralized exchange surely benefits from the transparent and permissionless nature of blockchain---anyone can scrutinize the underlying smart contract and pseudonymously participate. The use of liquidity pools brings about an unconventional trading mechanism where liquidity providers can act both as buyers and sellers. Fractional ownership of the pool is possible, and is represented with a special-purpose cryptocurrency called LP token. The pool also allows the contract to accommodate the demands from a large number of traders at the same time. Lastly, mathematical properties of CPMM guarantee that an exchange is possible regardless of the size of the order and the size of the pool.

The exchange system described, however, is not without drawbacks. The widely adopted CPMM equation $xy=k$ only permits \emph{one} trading pair per pool. In practice, a DEX will host several pools where one of the tokens in each pool is the chain's native currency. For example, a DEX on Ethereum blockchain will have several pools of the form X/ETH where X's are Ethereum tokens. An exchange from X to Y requires a ``hop'' X$\rightarrow$ETH$\rightarrow$Y and results in the trader paying transaction fee twice. Another inconvenience is that, to become a liquidity provider, one must deposit two tokens of equal value into the pool. Since users rarely possess idle cryptocurrencies of equal value, they may have to perform an unnecessary conversion so that they can maximize potential gain. Meanwhile, since the expected return from each pool is not the same, liquidity providers with several tokens have to make a difficult and strategic financial decision on how to convert and invest their funds.

Another interesting property of the CPMM equation $xy=k$ is that, when the price of the assets changes in an external market, an arbitrage opportunity exists so long as the total value of the two tokens in the pool is not equal. In other words, as liquidity providers deposit two tokens of equal value into the pool, this equilibrium will be maintained as traders use the pool to execute arbitrages. As a result, the liquidity pool can also be seen as an automated portfolio manager.
However, since liquidity providers may not necessarily want their portfolio balanced at the 50:50 ratio, this concept is expanded upon by the Balancer protocol where the token value ratio can be set to any proportion \cite{MM}. Nonetheless, the shortcoming is that isolated pools are needed to handle different ratios. While, in theory, each user can create a pool customized to their portfolio preference, doing so would leave the protocol with several shallow pools. Performing a swap over such an illiquid pool results in a high price impact, which is disadvantageous to the trader \cite{W}.

In this paper, we propose a novel dynamic exponent market maker (DEMM) that operates in conjunction with token-specific LP tokens. The equation underlying the proposed scheme can be seen as a generalization of those from Uniswap and Balancer. The features of the scheme can be summarized as follows.
\begin{itemize}
\item Liquidity providers are allowed to supply as many tokens as they wish into a single conglomerate pool. The exponent of the DEMM function is elastic and will automatically adjust whenever there is a provision or withdrawal. The advantages from doing so are fourfold. First, it greatly simplifies trading since any exchange can be made directly within the pool. Second, from liquidity providers' perspective, this is beneficial over the classical CPMM equation $xy=k$ in that incoming deposits are no longer restricted to be in the 50:50 ratio. Third, the pool functions as a \emph{personalized} portfolio manager. It automatically rebalances the allocation of assets so that the proportion of their net value stays the same. This is done for every user simultaneously. Finally, in comparison with the Balancer protocol, only a single pool is needed to accommodate each liquidity provider's portfolio preference. It can be seen as if pools in the aforementioned DEXs are combined to form a deeper pool under the proposed protocol. This benefits traders as a deep pool suffers less price impact compared to a shallow one \cite{W}.
\item Our approach enables native one-sided deposit and withdrawal, and is different from Bancor \cite{B1,B2} in that the protocol does not need to mint and manage an intermediate token. Partial liquidity providing may also result in a strange phenomenon named ``impermanent gain''. This is the opposite of impermanent loss as partial liquidity providers could gain more of every deposited token. In addition, new tokens can be added to the pool on the fly.
\item The proposed generalization differs from \cite{FK} in that it does not involve linear combinations of constant sum and constant product function. Rather, the dynamic invariant function results purely from manipulating the exponent of the constant product function. Additionally, in contrast to \cite{KFG}, our approach does not rely on an oracle, thus eliminating the risk of oracle manipulation or failure. 
\item It is possible to exploit the proposed protocol using flash loan attacks, and a preemptive countermeasure is necessary.
\end{itemize}

The remainder of this paper is organized as follows. This section ends with a brief summary of terminology. Section 2 provides some background on constant product market maker. The proposed protocol is outlined in Section 3, and Section 4 discusses some of its properties. The paper concludes in Section 5.

\textbf{Terminology and Notation.} In this paper, we use the term \emph{token} to refer to any cryptocurrency. Given $n$ ordered tokens, token $1,2,\ldots,n$, we say $\mathbf{r}=(r_1,r_2,\ldots,r_n)$ tokens to mean $r_1$ token 1, $r_2$ token 2, $\ldots$, and $r_n$ token $n$ collectively. They are said to have \emph{weight} $\mathbf{w}=(w_1,w_2,\ldots,w_n)$ if the ratio of the total market value of these tokens is 
\[w_1:w_2:\cdots:w_n\]
where $w_t>0$ for $t=1,2,\ldots,n$. For $\mathbf{x}=(x_1,x_2,\ldots,x_n)$ and $\mathbf{w}=(w_1,w_2,\ldots,w_n)$,
\[\mathbf{x}^\mathbf{w}=x_1^{w_1}x_2^{w_2}\cdots x_n^{w_n}.\]


\section{Constant Product Market Maker}


The functionality of an automated market maker can be formally described as a state machine, and we outline the general form of constant product market maker (CPMM) used by Balancer \cite{MM} in this section. Here, the protocol consists of a pool, which rests at a \emph{state}. The state of the pool can be changed via a state transition function, which is an external action or \emph{operation}.

Consider a liquidity pool with $n$ tokens, token $1,2,\ldots,n$, and weight $\mathbf{w}=(w_1,w_2,\ldots,w_n)$ where $w_t>0$ is the weight of token $t$, $t\in\{1,2,\ldots,n\}$. These tokens and their weight are preset at launch and can never be changed. Define an invariant function $f:\mathbb{R}^n\rightarrow\mathbb{R}$ by
\[f(\mathbf{x})=\mathbf{x}^\mathbf{w}=x_1^{w_1}x_2^{w_2}\cdots x_n^{w_n}.\]
Motivated by \cite{XPCF}, we describe the state of a liquidity pool under CPMM as an ordered pair
\[\chi=(\mathbf{r},L)\]
where $\mathbf{r}=(r_1,r_2,\ldots,r_n)$ represents the numbers of tokens in the pool and $L$ is the number of liquidity provider tokens, or LP tokens, in circulation. For simplicity we assume the protocol takes no transaction fee, and we ignore gas fee.

To initiate a pool, the following operation is called.

\begin{operation}[Initialize]
The genesis liquidity provider supplies token $1,2,\ldots,n$ with weight $\mathbf{w}$ and receives a set number of LP tokens as a receipt.
\end{operation}

There are three ways to interact with the pool: trade, provide liquidity, and withdraw liquidity. We let $\chi=(\mathbf{r},L)$ be the current state of the pool, and describe the three operations as follows.

\begin{operation}[Trade]\label{o:trade1}
A trader may exchange $\partial r_i$ token $i$ for $\partial r_o$ token $o$ where
\begin{equation}\label{e:trade1}
\partial r_o=r_o-r_o\left(\frac{r_i}{r_i+\partial r_i}\right)^\frac{w_i}{w_o}.
\end{equation}
This changes the state of the pool to
\[\chi'=(\mathbf{r'},L)\]
where $r_i'=r_i+\partial r_i$, $r_o'=r_o-\partial r_o$, and $r_t'=r_t$ for $t\in\{1,2,\ldots,n\}\setminus\{i,o\}$.
\end{operation}

Note that equation \eqref{e:trade1} can be rearranged as
\[r_i^{w_i}r_o^{w_o}=(r_i+\partial r_i)^{w_i}(r_o-\partial r_o)^{w_o}.\]
It may be helpful to think of this trading mechanism in conjunction with the function $f(\mathbf{x})$. The ubiquitous terminologies \emph{constant product} and \emph{invariant function} come from the fact that Operation \ref{o:trade1} does not change the value of this function. In other words, $f(\mathbf{r})=f(\mathbf{r'})$, and trading simply moves the amount of tokens within the pool along the surface of $f(\mathbf{x})$. When $n=2$ and $\mathbf{w}=(1,1)$, this condition reduces to the famous $xy=k$ formula used by Uniswap and many other DEXs.

Meanwhile, the LP token represents a proof of fractional ownership of the $\mathbf{r}$ tokens within the pool and is practically a certificate of deposit or a receipt. A deposit and withdrawal of tokens has to be done in proportion to $\mathbf{r}$.

\begin{operation}[Provide liquidity]
A liquidity provider may deposit $\alpha\mathbf{r}$ tokens and receive $\alpha L$ newly minted LP tokens for some $\alpha>0$. The state of the pool becomes
\[\chi'=((1+\alpha)\mathbf{r},(1+\alpha)L).\]
\end{operation}

\begin{operation}[Withdraw liquidity]
An existing liquidity provider may redeem $\alpha L$ LP tokens for some $0<\alpha\leq1$ for his or her share of the pool. The $\alpha L$ LP tokens redeemed is burned, and the pool is left with the state
\[\chi'=((1-\alpha)\mathbf{r},(1-\alpha)L).\]
\end{operation}

It can be shown \cite{MM} that the relative spot price between any two tokens, say token $o$ and token $i$, can be given as
\begin{equation}\label{e:price}
P_{o/i}=\frac{\frac{r_i}{w_i}}{\frac{r_o}{w_o}}.
\end{equation}
Note that this ratio is unaffected by liquidity provision and withdrawal, and can only be changed with trading. For any pair of tokens where this price does not equal that of an external market, an arbitrage opportunity occurs. Thus, in a rational market, the numbers of token $1,2,\ldots,n$ in the pool move in accordance with the market so as to correct the spot price. Next, we restate the standard feature of CPMM via an auxiliary definition.

\begin{definition}
An automated market maker with an invariant function $f(\mathbf{x})=\mathbf{x}^\mathbf{w}$ is \emph{exponent-balanced} if $\mathbf{w}$ is the weight of tokens in the pool.
\end{definition}

\begin{proposition}\cite{MM}\label{p:eb1}
Constant product market maker is exponent-balanced, given there is no arbitrage opportunity.
\end{proposition}

Since the weight of tokens in the pool is kept at $\mathbf{w}$ by arbitragers, liquidity provision and withdrawal must follow. In other words, Proposition \ref{p:eb1} implies that liquidity pool doubles as an automated portfolio manager. We illustrate some of the concepts discussed in the following example.

\begin{example}\label{ex:cpmm}
Alice would like to set up a CPMM pool with two tokens, token $s$ and token $t$, and weight $\mathbf{w}=(1,1)$. The invariant function is then $f(x_s,x_t)=x_sx_t$. Suppose that the unit price of token $s$ is \$2, and the unit price of token $t$ is \$10. Alice initializes the pool with $(20,4)$ tokens, i.e., 20 token $s$ and 4 token $t$, and receives 1 LP token. The state of the pool is $\chi=((20,4),1)$.

Now, Bob, who owns 12 token $t$, would like to become a liquidity provider for this pool. Since the pool only admits deposits that are proportionate to its current state, Bob needs to additionally acquire 60 token $s$ so that he can supply $(60,12)$ tokens into the pool. Bob receives 3 LP tokens, and the state of the pool becomes $\chi'=((80,16),4)$. Here, it can be seen as if 1 LP token represents 20 token $s$ and 4 token $t$.

Suppose that a trader would like to exchange 1 token $s$ for token $t$. Had the exchange been done while the pool was at $\chi=((20,4),1)$, he or she would have received $4-4\left(\frac{20}{20+1}\right)\approx0.1905$ token $t$. This agrees with the constant product constraint since $20\cdot4=80\approx(20+1)\cdot(4-0.1905)$. However, if the same trader executes the trade at the current state $\chi'$, he or she will receive $16-16\left(\frac{80}{80+1}\right)\approx0.1975$ token $t$, which once again agrees with the constant product constraint as $80\cdot16=1280\approx(80+1)\cdot(16-0.1975)$. Here, note that deeper pool results in better exchange rate.

Suppose now that the unit prices of token $s$ and $t$ in an external market become \$64 and \$5 respectively. It is not hard to see that an arbitrager can buy the cheaper token $t$ from another market, use the pool to swap for token $s$, and sell token $s$ in an external market for profit. The pool will eventually reach a new equilibrium of $((10,128),4)$. Here, the weight of the tokens in the pool is $10\cdot64:128\cdot5=1:1$. Among $\mathbf{r}=(10,128)$ tokens in the pool, Alice's fractional share is $\frac{1}{4}(10,128)=(2.5,32)$ while Bob is entitled to $\frac{3}{4}(10,128)=(7.5,96)$ tokens. After the change in price, 1 LP token now represents 2.5 token $s$ and 32 token $t$.
\end{example}

\section{Dynamic Exponent Market Maker}

We see from the previous section that the exponent of the invariant function corresponds to the weight of tokens within the pool. This weight, which is fixed for each pool and can never be changed, dictates the proportion in which liquidity providers make a deposit. If we were to permit a free-form deposit, one way to do that is to allow the exponent of the invariant function to fluctuate based on incoming funds from liquidity providers. The adjustment here has to be in such a way that it complies with Proposition \ref{p:eb1}, or else there will be an immediate arbitrage opportunity.

Another complication that arises is the representation of pool shares. Under CPMM, this is facilitated by the use of LP token, which serves to represent fractional ownership of every token within the pool. This one-token-fit-all LP token is no longer viable since the weight of tokens in the pool will shift from time to time, and one possible solution is to instead issue a separate LP token for each asset. These intuitions result in the proposed dynamic exponent market maker (DEMM), an automated market maker with dynamic exponent and token-specific LP token, which we shall make precise in this section.

Under DEMM, the exponent of the invariant function is allowed to change and depends on incoming funds from liquidity providers. To accommodate deposits of varying proportion, LP tokens are issued for \emph{each} token and represent fractional ownership of that particular token rather than all tokens across the pool. As we will see, while in operation, the protocol maintains that the \emph{exponent}, the \emph{weight}, and the \emph{numbers of LP tokens in circulation} are equal.

Denote the state of a DEMM liquidity pool with token $1,2,\ldots,n$ as an ordered pair
\[\chi=(\mathbf{r},\mathbf{w})\]
where $\mathbf{r}=(r_1,r_2,\ldots,r_n)$ is the numbers of tokens in the pool. At this state, the invariant function is defined as
\[f(\mathbf{x})=\mathbf{x}^\mathbf{w}=x_1^{w_1}x_2^{w_2}\cdots x_n^{w_n}.\]
Note that, unlike the predecessor, this function is linked to the state of the pool and is not predetermined.

The pool is set up by the following operation.

\begin{operation}[Initialize]\label{o:initialize2}
The genesis liquidity provider forms a pool with $\mathbf{r}$ tokens of equal value. The state of the pool is initialized to $\chi=(\mathbf{r},\mathbf{1})$ where $\mathbf{1}$ is the all-ones vector. The genesis liquidity provider receives $1$ LP token $t$ for each $t=1,2,\ldots,n$.
\end{operation}

Note that a counterpart LP token exists for every token in the pool, and the initial invariant function is $f(\mathbf{x})=x_1x_2\cdots x_n$. Next, we describe three ways to interact with the pool: trade, provide liquidity, and withdraw liquidity. Let $\chi=(\mathbf{r},\mathbf{w})$ be the current state of the pool, where the entries of $\mathbf{r}$ and $\mathbf{w}$ are strictly positive. This means the present invariant function is $f(\mathbf{x})=x_1^{w_1}x_2^{w_2}\cdots x_n^{w_n}$. We also remark here that trading under DEMM is identical to that of CPMM and does not affect the invariant function.

\begin{operation}[Trade]\label{o:trade2}
A user can swap $\partial r_i$ token $i$ for
\begin{equation}\label{e:trade2}
\partial r_o=r_o-r_o\left(\frac{r_i}{r_i+\partial r_i}\right)^\frac{w_i}{w_o}.
\end{equation}
token $o$. The updated pool is $\chi'=(\mathbf{r'},\mathbf{w})$ where $r_i'=r_i+\partial r_i$, $r_o'=r_o-\partial r_o$, and $r_t'=r_t$ for $t\in\{1,2,\ldots,n\}\setminus\{i,o\}$.
\end{operation}

\begin{operation}[Provide liquidity]\label{o:provide2}
Suppose that a liquidity provider would like to deposit $\mathbf{\Delta r}$ tokens. This will change the state of the pool to $\chi'=(\mathbf{r'},\mathbf{w'})$ where
\[\mathbf{r'}=\mathbf{r}+\mathbf{\Delta r}\]
and
\begin{equation}\label{e:provide2}
\mathbf{w'}=\left(\left(\frac{r_1+\Delta r_1}{r_1}\right)w_1,\left(\frac{r_2+\Delta r_2}{r_2}\right)w_2,\ldots,\left(\frac{r_n+\Delta r_n}{r_n}\right)w_n\right).
\end{equation}
For each $t\in\{1,2,\ldots,n\}$, the liquidity provider will receive $\frac{\Delta r_t}{r_t}w_t$ LP token $t$.
\end{operation}

\begin{operation}[Withdraw liquidity]\label{o:withdraw2}
A redemption of $\mathbf{\Delta w}$ LP tokens changes the state of the pool to $\chi'=(\mathbf{r'},\mathbf{w'})$ where
\[\mathbf{r'}=\left(r_1\left(\frac{w_1-\Delta w_1}{w_1}\right),r_2\left(\frac{w_2-\Delta w_2}{w_2}\right),\ldots,r_n\left(\frac{w_n-\Delta w_n}{w_n}\right)\right)\]
and
\begin{equation}\label{e:withdraw2}
\mathbf{w'}=\mathbf{w}-\mathbf{\Delta w}.
\end{equation}
In other words, the withdrawer receives $\left(r_1\frac{\Delta w_1}{w_1},r_2\frac{\Delta w_2}{w_2},\ldots,r_n\frac{\Delta w_n}{w_n}\right)$ token.
\end{operation}

Intuitively, token-specific LP token represents fractional ownership of its respective asset. Particularly, the proportion by which a user holds an LP token $t$ relative to all LP token $t$ in circulation is precisely his or her share of token $t$ in the pool. During Operation \ref{o:provide2}, a liquidity provider does not have to provide every asset. That is, $\Delta r_t$ is allowed to be zero, and $\Delta r_t=0$ simply means that the protocol will not update the $t$\textsuperscript{th} coordinate of $\mathbf{r}$ and $\mathbf{w}$, and the liquidity provider receives no LP token $t$. Similarly, liquidity withdrawal with $\Delta w_t=0$ means token $t$, hence the $t$\textsuperscript{th} coordinate, is unaffected. In the following proposition, we prove that the numbers of LP tokens is tied to the parameter $\mathbf{w}$ of the pool.

\begin{proposition}
If the state of the pool is $\chi=(\mathbf{r},\mathbf{w})$, then $\mathbf{w}$ is the numbers of LP tokens in circulation.
\end{proposition}

\begin{proof}
It is clear that the statement holds after the pool is initialized via Operation \ref{o:initialize2}. Next, we suppose that the statement holds, and we will show that it remains so after any of the other three operations is performed.

Since Operation \ref{o:trade2} does not change $\mathbf{w}$ nor the numbers of LP tokens in circulation, the statement trivially holds after the pool is being used for trading.

During Operation \ref{o:provide2}, the liquidity provider receives $\frac{\Delta r_t}{r_t}w_t$ LP token $t$ for each $t\in\{1,2,\ldots,n\}$. Thus, the number of LP token $t$ in circulation is
\[w_t+\frac{\Delta r_t}{r_t}w_t=\left(\frac{r_t+\Delta r_t}{r_t}\right)w_t=w_t'\]
for each $t$, and $\mathbf{w'}$ is then the current numbers of LP tokens in circulation.

Finally, a redemption of $\mathbf{\Delta w}$ LP tokens for liquidity withdrawal results in an update $\mathbf{w'}=\mathbf{w}-\mathbf{\Delta w}$ as given in \eqref{e:withdraw2}. This completes the proof of the proposition.
\end{proof}

We have seen that the use of individualized LP token by DEMM allows liquidity providers to deposit and withdraw assets in any proportion. However, to accommodate the fluctuation of the ratio of the total value of each token in the pool, the exponent of the invariant function needs to adapt. In other words, the dynamic exponent of DEMM has to correspond precisely to the weight of tokens within the pool. This will be proven as the next result, which is a direct generalization of Proposition \ref{p:eb1}.


\begin{proposition}\label{p:balance}
Dynamic exponent market maker is exponent-balanced, given there is no arbitrage opportunity.
\end{proposition}

\begin{proof}
In the beginning, the pool is exponent-balanced since the genesis liquidity provider deposits tokens of equal value into the pool, and the initial exponent is $\mathbf{w}=\mathbf{1}$. The invariant function remains unchanged as traders and arbitragers use Operation \ref{o:trade2} to swap tokens, and so it follows from Proposition \ref{p:eb1} that the pool remains exponent-balanced. We are left to show that liquidity provision and withdrawal preserve the required property. Let $s,t\in\{1,2,\ldots,n\}$, and suppose that the current state of the pool is $\chi=(\mathbf{r},\mathbf{w})$. This means the ratio of the total value of token $s$ to token $t$ in the pool is $w_s:w_t$.

Suppose now that a liquidity provider deposits $\mathbf{\Delta r}$ tokens into the pool. This changes the aforementioned ratio to
\begin{equation}\label{e:proof1}
\left(\frac{r_s+\Delta r_s}{r_s}\right)w_s:\left(\frac{r_t+\Delta r_t}{r_t}\right)w_t
\end{equation}
since $\frac{r_s+\Delta r_s}{r_s}$ and $\frac{r_t+\Delta r_t}{r_t}$ are the proportions by which the total value of tokens $s$ and $t$ grow respectively. We now see from \eqref{e:provide2} that the ratio \eqref{e:proof1} equals $w_s':w_t'$ after the deposit. This holds for any $s$ and $t$, and so the pool remains exponent-balanced.

On the other hand, suppose that a liquidity provider redeems $\mathbf{\Delta w}$ LP tokens. Again, the proportions by which the total value of tokens $s$ and $t$ shrink are $\frac{w_s-\Delta w_s}{w_s}$ and $\frac{w_t-\Delta w_t}{w_t}$ respectively. Thus, after the redemption, the ratio of the total value of token $s$ to token $t$ is
\[\left(\frac{w_s-\Delta w_s}{w_s}\right)w_s:\left(\frac{w_t-\Delta w_t}{w_t}\right)w_t=w_s-\Delta w_s:w_t-\Delta w_t=w_s':w_t',\]
and the pool once again remains exponent-balanced.
\end{proof}

In a way, constant product market maker can be seen as a restricted instance of our proposed scheme. A generic LP token under CPMM is a weight-specific concoction of specialized LP tokens. While Proposition \ref{p:eb1} establishes that the exponent $\mathbf{w}$ of the invariant function corresponds to the proportion of tokens by total value in the pool, CPMM does not allow this parameter to change. On the other hand, dynamic exponent market maker allows liquidity providing of any proportion, and it alters the exponent in such a way that the pool stays balanced after the deposit. Meanwhile, the impact on the exponent is reflected back to the liquidity provider in the form of LP tokens. Piecing this together, we relate the weight of the tokens deposited and withdrawn in the following theorem.


\begin{theorem}
Dynamic exponent market maker preserves the ratio of the total value of tokens at the time of deposit, given there is no arbitrage opportunity.
\end{theorem}

\begin{proof}
Let $s,t\in\{1,2,\ldots,n\}$. It suffices to show that the total value ratio between tokens $s$ and $t$ is the same at the time of liquidity provision and withdrawal.

Suppose that a liquidity provider supplies $\mathbf{\Delta r}$ tokens into a pool at the state $\chi=(\mathbf{r},\mathbf{w})$. It follows from Proposition \ref{p:balance} that the proportion of tokens $s$ and $t$ by net value in the pool is $w_s:w_t$, and so the ratio of the unit price of token $s$ to token $t$ is $\frac{w_s}{r_s}:\frac{w_t}{r_t}$. This means the liquidity provider deposits tokens $s$ and $t$ by the ratio
\[\frac{w_s}{r_s}\Delta r_s:\frac{w_t}{r_t}\Delta r_t\]
in total value. Finally, according to Operation \ref{o:provide2}, this person will receive $\frac{\Delta r_s}{r_s}w_s$ LP token $s$ and $\frac{\Delta r_t}{r_t}w_t$ LP token $t$.

Suppose now that the same liquidity provider withdraws the fund some time later when the state of the pool is $\chi'=(\mathbf{r'},\mathbf{w'})$. By the same argument, the ratio of the current unit price of token $s$ to token $t$ is $\frac{w_s'}{r_s'}:\frac{w_t'}{r_t'}$. Given that $\frac{\Delta r_s}{r_s}w_s$ LP token $s$ and $\frac{\Delta r_t}{r_t}w_t$ LP token $t$ is redeemed, the numbers of tokens $s$ and $t$ this liquidity provider receives is
\[r_s'\frac{\frac{\Delta r_s}{r_s}w_s}{w_s'}\quad\textrm{and}\quad r_t'\frac{\frac{\Delta r_t}{r_t}w_t}{w_t'}\]
respectively. Thus, the ratio of the total value of token $s$ and token $t$ received is
\[r_s'\frac{\frac{\Delta r_s}{r_s}w_s}{w_s'}\cdot\frac{w_s'}{r_s'}:r_t'\frac{\frac{\Delta r_t}{r_t}w_t}{w_t'}\cdot\frac{w_t'}{r_t'}=\frac{w_s}{r_s}\Delta r_s:\frac{w_t}{r_t}\Delta r_t\]
as required.
\end{proof}

It can be seen that each liquidity provider's assortment of LP tokens represents his or her portfolio preference. These LP tokens add up to the exponent of the invariant function, which now represents the coalition of all preferences. We shall illustrate this concept in the next example.

\begin{figure}
\begin{center}
\includegraphics{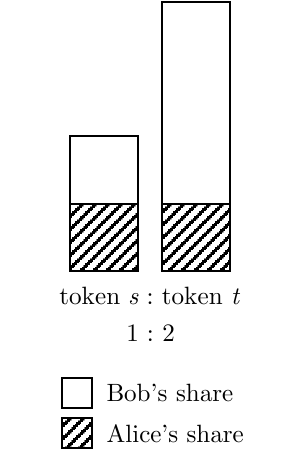}
\caption{\label{f:chart} A visual representation of the pool from Example \ref{ex:pool} and its allocation of value}
\end{center}
\end{figure}

\begin{example}\label{ex:pool}
Suppose that the unit price of tokens $s$ and $t$ is \$2 and \$10 respectively. Alice initiates a DEMM pool with 20 token $s$ and 4 token $t$, and receives 1 LP token $s$ and 1 LP token $t$. Here, the state of the pool is $\chi=((20,4),(1,1))$, and the invariant function is $f(x_s,x_t)=x_sx_t$.

Meanwhile, if Bob deposits 20 token $s$ and 12 token $t$ into the same pool, then he will receive 1 LP token $s$ and 3 LP token $t$. This updates the state of the pool to $((40,16),(2,4))$ and makes $f(x_s,x_t)=x_s^2x_t^4$ the invariant function.

Given the current exponent $\mathbf{w}=(2,4)$, the pool will maintain the weight between tokens $s$ and $t$ to be $2:4=1:2$. Among all tokens in the pool, Alice is entitled to half of token $s$ and one-fourth of token $t$, and so her personal weight is $\frac{1}{2}\cdot1:\frac{1}{4}\cdot2=1:1$. This corresponds to the fact that she deposits tokens $s$ and $t$ of equal value into the pool, and she holds 1 LP token $s$ and 1 LP token $t$. Bob, on the other hand, is entitled to half of token $s$ and three-fourth of token $t$. The weight of tokens $s$ and $t$ in his holding is $\frac{1}{2}\cdot1:\frac{3}{4}\cdot2=1:3$. Again, this corresponds to the fact that he deposits \$40 worth of token $s$ and \$120 worth of token $t$ into the pool, and he holds 1 LP token $s$ and 3 LP token $t$. Alice and Bob's share of the pool is illustrated in Figure \ref{f:chart}.

Suppose, for example, that the unit price of tokens $s$ and $t$ has become \$64 and \$5 respectively. The state of the pool after arbitraging is $((2.5,64),(2,4))$. Among these $\mathbf{r}=(2.5,64)$ token, Alice is entitled to 1.25 token $s$ and 16 token $t$, and they are worth \$80 and \$80 respectively. Bob is entitled to 1.25 token $s$ and 48 token $t$, and they are worth \$80 and \$240 respectively.
\end{example}

\section{Risks and Implementation Issues}

In this section, we discuss some known issues and additional features of dynamic exponent market maker.

We first remark that the value of the invariant function $f(\mathbf{x})=x_1^{w_1}x_2^{w_2}\cdots x_n^{w_n}$ could easily overflow since the exponent corresponds to the numbers of LP tokens. This value, however, is never explicitly needed. Trading is executed in accordance with \eqref{e:trade2}, and only an update on the status $\chi=(\mathbf{r},\mathbf{w})$ of the pool is necessary during liquidity provision and withdrawal.

As we have noted, one inconvenience of constant product market maker is that prospective liquidity providers need to have tokens by the right weight. Partial (one-sided) liquidity providing can be done by preceding it with a token swap \cite{MM}, but this is suboptimal since the pool may not be in equilibrium by the time of deposit. Alternatively, the protocol can lend or mint a counterpart token in matching value \cite{B1,B2,AF}, and the downside here is that potential profit is reduced.

In our proposed protocol, users may make a deposit in any amount, meaning that a true and natural partial liquidity providing is viable. Specifically, since the elements of $\Delta\mathbf{r}$ from Operation \ref{o:provide2} can be zero, liquidity providers can choose to only provide tokens that they wish. Additionally, since the LP token for each asset is treated separately, users also have the option to redeem only some specific assets and leave the rest in the pool.

\subsection{Price Impact from Liquidity Provision}

By design, the exponent of the invariant function under DEMM is automatically adjusted in proportion to incoming funds from liquidity providers. While they can supply (or withdraw) tokens in any quantity, such actions will not result in an arbitrage opportunity. This will be proven in the next proposition.

\begin{proposition}
Relative spot price between any two tokens is unchanged after liquidity provision and withdrawal.
\end{proposition}

\begin{proof}
Let $\chi=(\mathbf{r},\mathbf{w})$ and $\chi'=(\mathbf{r'},\mathbf{w'})$ be the states of the pool before and after liquidity provision. From \eqref{e:price}, the spot price of token $o$ relative to token $i$ at state $\chi$ is given as
\[P_{o/i}=\frac{\frac{r_i}{w_i}}{\frac{r_o}{w_o}}.\]
After liquidity provision, this quotient remains fixed since
\[P'_{o/i}=\frac{\frac{r'_i}{w'_i}}{\frac{r'_o}{w'_o}}=\frac{\frac{r_i+\Delta r_i}{\left(\frac{r_i+\Delta r_i}{r_i}\right)w_i}}{\frac{r_o+\Delta r_o}{\left(\frac{r_o+\Delta r_o}{r_o}\right)w_o}}=\frac{\frac{r_i}{w_i}}{\frac{r_o}{w_o}}.\]
The statement also holds for liquidity withdrawal since it is simply the reverse of the above process.
\end{proof}

This proposition means that, for an infinitesimal exchange, relative token price is unaffected by liquidity provision. However, the assertion will not hold for a larger regular trade. A particularly curious scenario arises when a liquidity provider supplies tokens partially. Here, there are contradicting principles, one being that an asset should depreciate as its liquidity grows, and the other being that deeper pool should result in a better exchange rate. We will see in the next theorem that the second speculation always overrules the first one in the proposed scheme.

\begin{theorem}\label{t:rate}
Under DEMM, liquidity provision can only benefit traders. In other words, a swap results in better (or the same) outcome as liquidity providers supply funds.
\end{theorem}

\begin{proof}
We will show that liquidity providing one single asset improves the exchange rate of that asset in \emph{both} buying and selling directions. Applying the argument successively for every deposited token yields the desired result.

Let $s,t\in\{1,2,\ldots,n\}$. Suppose that $\Delta r_s$ token $s$ is being supplied into the pool so that its state changes from $\chi=(\mathbf{r},\mathbf{w})$ to $\chi'=(\mathbf{r'},\mathbf{w'})$. Let $\alpha=\frac{r_s+\Delta r_s}{r_s}>1$, so we can write $r_s'=\alpha r_s$ and $w_s'=\alpha w_s$. We state the following claim, which can be proved using standard techniques from calculus.

\begin{quote}
\textbf{Claim} \begin{tabular}[t]{@{}l}
For $0<A$ and $1<x$, $\left(1+\frac{A}{x}\right)^x>1+A$. \\
For $0<A<1$ and $1<x$, $\left(1-\frac{A}{x}\right)^x>1-A$.
\end{tabular}
\end{quote}

Suppose that $\partial r_s$ token $s$ can be exchanged for $\partial r_t$ token $t$ at state $\chi$; that is,
\begin{align*}
(r_s+\partial r_s)^{w_s}(r_t-\partial r_t)^{w_t} & =r_s^{w_s}r_t^{w_t}, \\
\intertext{which is equivalent to}
\left(\frac{r_s+\partial r_s}{r_s}\right)^{w_s} & =\left(\frac{r_t}{r_t-\partial r_t}\right)^{w_t}.
\end{align*}
We will show that the same exchange at state $\chi'$ increases the value of the invariant. Starting from the claim,
\begin{align*}
\left(1+\frac{\frac{\partial r_s}{r_s}}{\alpha}\right)^\alpha & >1+\frac{\partial r_s}{r_s} \\
\left(\frac{\alpha r_s+\partial r_s}{\alpha r_s}\right)^{\alpha w_s} & >\left(\frac{r_s+\partial r_s}{r_s}\right)^{w_s} \\
\left(\frac{r_s'+\partial r_s}{r_s'}\right)^{w_s'} & >\left(\frac{r_t}{r_t-\partial r_t}\right)^{w_t} \\
(r_s'+\partial r_s)^{w_s'}(r_t-\partial r_t)^{w_t} & >r_s'^{w_s'}r_t^{w_t}.
\end{align*}
This means the same trade is admissible at state $\chi'$, and it is impossible that the exchange rate worsens.

On the other hand, suppose now that one can trade $\widehat{\partial r_t}$ token $t$ for $\widehat{\partial r_s}$ token $s$ at state $\chi$. We now have
\begin{align*}
\left(r_s-\widehat{\partial r_s}\right)^{w_s}\left(r_t+\widehat{\partial r_t}\right)^{w_t} & =r_s^{w_s}r_t^{w_t}, \\
\intertext{or}
\left(\frac{r_s-\widehat{\partial r_s}}{r_s}\right)^{w_s} & =\left(\frac{r_t}{r_t+\widehat{\partial r_t}}\right)^{w_t}.
\end{align*}
Following the same line or argument,
\begin{align*}
\left(1-\frac{\frac{\widehat{\partial r_s}}{r_s}}{\alpha}\right)^\alpha & >1-\frac{\widehat{\partial r_s}}{r_s} \\
\left(\frac{\alpha r_s-\widehat{\partial r_s}}{\alpha r_s}\right)^{\alpha w_s} & >\left(\frac{r_s-\widehat{\partial r_s}}{r_s}\right)^{w_s} \\
\left(\frac{r_s'-\widehat{\partial r_s}}{r_s'}\right)^{w_s'} & >\left(\frac{r_t}{r_t+\widehat{\partial r_t}}\right)^{w_t} \\
\left(r_s'-\widehat{\partial r_s}\right)^{w_s'}\left(r_t+\widehat{\partial r_t}\right)^{w_t} & >r_s'^{w_s'}r_t^{w_t}.
\end{align*}
Again, one can commit the same trade, and this completes the proof of the theorem.
\end{proof}

While Theorem \ref{t:rate} may seem counterintuitive from a certain perspective, it is coherent with the fact that liquidity providing also increases the exponent of the invariant function. Loosely speaking, as the function $f(\mathbf{x})=\mathbf{x}^\mathbf{w}$ becomes more sensitive to the change in $\mathbf{x}$, it takes less input tokens to execute the same trade. As thus, with respect to the axiomatic framework of AMM set up in \cite{BF}, our work does not comply with the monotone in liquidity (PML) property.

Note that the converse of Theorem \ref{t:rate} also holds: withdrawing liquidity can only drive the exchange rate lower. We now finish this subsection with an example.

\begin{example}\label{ex:rate}
Suppose that the unit price of tokens $s$ and $t$ is \$2 and \$10 respectively. Carol sets up a DEMM pool with 40 token $s$ and 8 token $t$. This makes $\chi=((40,8),(1,1))$ the initial state of the pool. Later, Dave one-sidedly provides 8 token $t$, shifting the pool to $\chi'=((40,16),(1,2))$. Figure \ref{f:rate} plots the amount of token $t$ received from an exchange of varying volumes of token $s$ (and vice versa) at the market price and under DEMM at the two states. Note that the pool at state $\chi$ is simply a classical CPMM.

\begin{figure}
\begin{center}
\includegraphics{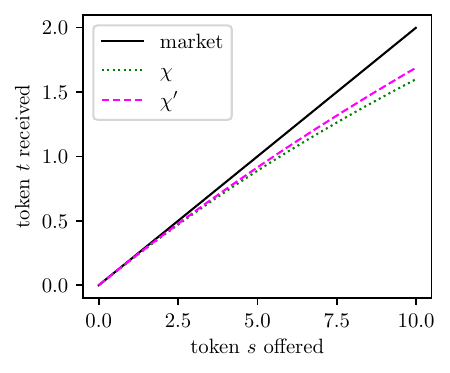}\quad\includegraphics{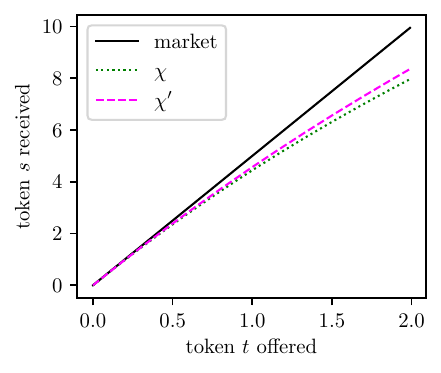}
\caption{\label{f:rate} Token $t$ received from trading away token $s$ (left) and token $s$ received from trading away token $t$ (right) from Example \ref{ex:rate}}
\end{center}
\end{figure}

\end{example}

\subsection{Transaction Fee}


In practice, traders on DEXs will normally be charged a transaction fee, and we briefly discuss its mechanism here. At first glance, it may seem necessary to decide whether to tax incoming token, outgoing token, or both during a swap. These options, however, are somewhat equivalent. Suppose that the protocol charges $(1-\rho)100\%$ transaction fee, then the amount of token received from the trade operation, i.e., $\partial r_o$ from \eqref{e:trade2}, is given instead by
\[\partial r_o=r_o-r_o\left(\frac{r_i}{r_i+\rho\partial r_i}\right)^\frac{w_i}{w_o}.\]
In other words, the exchange is executed as if the trader only puts in $\rho\partial r_i$ token $i$. Here, the unaccounted $(1-\rho)\partial r_i$ token $i$ serves as a transaction fee, and it is distributed among all LP token $i$ holders by the proportion of their holding. Alternatively, one may regard that $\partial r_i$ token $i$ in its entirety has entered the pool, but it withholds
\[r_o\left(\frac{r_i}{r_i+\rho\partial r_i}\right)^\frac{w_i}{w_o}-r_o\left(\frac{r_i}{r_i+\partial r_i}\right)^\frac{w_i}{w_o}\]
token $o$ that the trader would have otherwise received. Again, the token retained here plays the role of a transaction fee, and is allocated to the liquidity providers. Note that a liquidity provider benefits from a trade only when it directly involves one of the tokens he or she has supplied, and this system incentivizes users to supply and maintain frequently traded tokens in the pool.

\subsection{Impermanent Loss}

Impermanent loss, also called divergence loss, is a risk that comes with liquidity providing for decentralized exchanges and occurs when relative token price changes. We refer interested readers to \cite{TPL} for a comprehensive survey on the topic. Since DEMM preserves the weight of the tokens at the time of deposit, liquidity providers suffer impermanent loss to some extent. This concept is demonstrated in the next example.

\begin{figure}
\begin{center}
\begin{picture}(370,160)
\put(15,150){token $s$ = \$2 and token $t$ = \$10}
\put(228,150){token $s$ = \$64 and token $t$ = \$5}
\put(0,70){Alice}
\put(35,70){\begin{tabular}{|c|c|}
\hline
20 token $s$ & 4 token $t$ \\
{[}\$40{]} & {[}\$40{]} \\
\hline
\end{tabular}}
\put(235,120){\begin{tabular}{|c|c|}
\hline
20 token $s$ & 4 token $t$ \\
{[}\$1280{]} & {[}\$20{]} \\
\hline
\end{tabular}}
\put(235,70){\begin{tabular}{|c|c|}
\hline
$2.5$ token $s$ & $32$ token $t$ \\
{[}\$160{]} & {[}\$160{]} \\
\hline
\end{tabular}}
\put(235,20){\begin{tabular}{|c|c|}
\hline
1.25 token $s$ & 16 token $t$ \\
{[}\$80{]} & {[}\$80{]} \\
\hline
\end{tabular}}
\put(159,72){\vector(3,2){75}}
\put(159,72){\vector(1,0){75}}
\put(159,72){\vector(3,-2){75}}
\put(182,105){hold}
\put(182,75){CPMM}
\put(193,62){1:1}
\put(159,35){DEMM}
\end{picture} \\
\begin{picture}(370,160)
\put(0,70){Bob}
\put(30,70){\begin{tabular}{|c|c|}
\hline
20 token $s$ & 12 token $t$ \\
{[}\$40{]} & {[}\$120{]} \\
\hline
\end{tabular}}
\put(235,120){\begin{tabular}{|c|c|}
\hline
20 token $s$ & 12 token $t$ \\
{[}\$1280{]} & {[}\$60{]} \\
\hline
\end{tabular}}
\put(235,70){\begin{tabular}{|c|c|}
\hline
$\frac{5}{4\sqrt{2}}$ token $s$ & $24\sqrt{2}$ token $t$ \\
{[}$\approx\$56.57${]} & {[}$\approx\$169.71${]} \\
\hline
\end{tabular}}
\put(235,20){\begin{tabular}{|c|c|}
\hline
1.25 token $s$ & 48 token $t$ \\
{[}\$80{]} & {[}\$240{]} \\
\hline
\end{tabular}}
\put(159,72){\vector(3,2){75}}
\put(159,72){\vector(1,0){75}}
\put(159,72){\vector(3,-2){75}}
\put(182,105){hold}
\put(182,75){CPMM}
\put(193,62){1:3}
\put(160,35){DEMM}
\end{picture}
\caption{\label{f:loss} Holding options and impermanent loss for Alice (top) and Bob (bottom) from Example \ref{ex:loss}}
\end{center}
\end{figure}
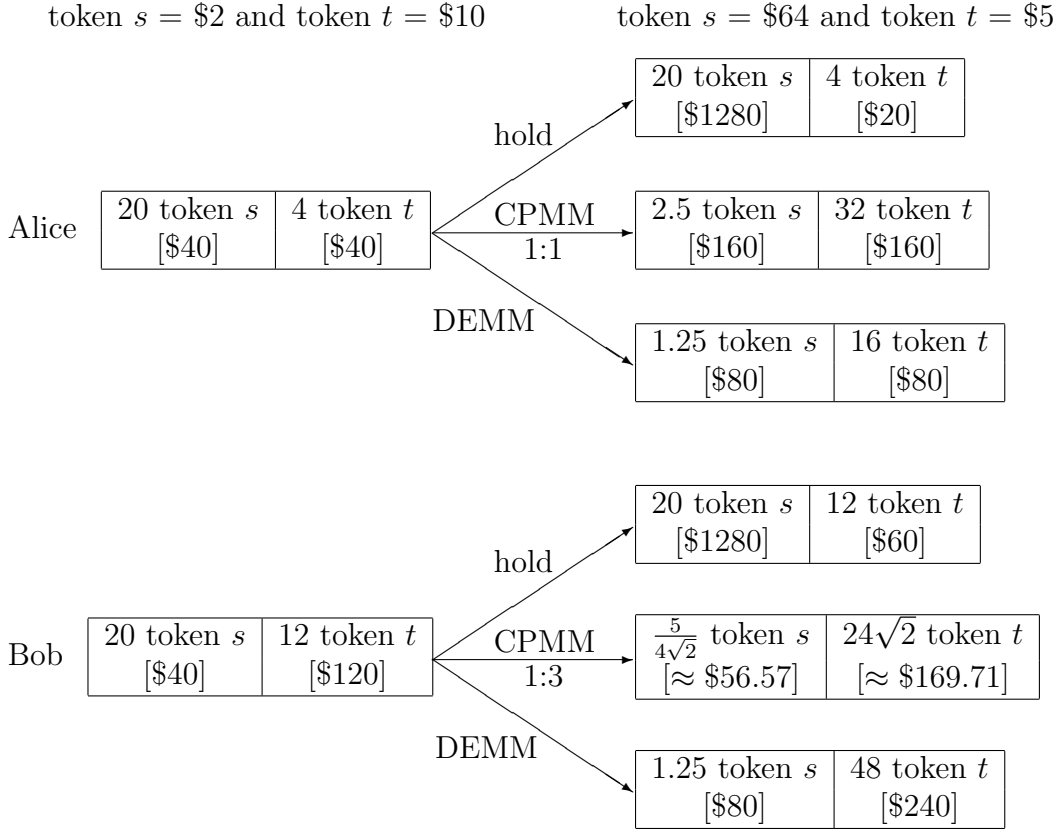

\begin{example}\label{ex:loss}
We follow up with Alice who has 20 token $s$ and 4 token $t$ from Examples \ref{ex:cpmm} and \ref{ex:pool} and consider her three options: hold, provide liquidity for CPMM, and provide liquidity for DEMM. The results are outlined in Figure \ref{f:loss} (top). While Alice's net worth in dollars increases across all scenarios, her best option is to simply hold onto the tokens she starts off with.

Now, consider Bob who owns 20 token $s$ and 12 token $t$. His situation is portrayed in Figure \ref{f:loss} (below). The weight of his initial holding is $1:3$, and he could have supplied his fund into a CPMM pool with that exact weight. It can be shown that, if the unit prices of tokens $s$ and $t$ move to \$64 and \$5 respectively, then he will end up with $\frac{5}{4\sqrt{2}}$ token $s$ and $24\sqrt{2}$ token $t$. On the other hand, we see from Example \ref{ex:pool} that, had Bob joined Alice's pool in DEMM, he would have had $(1.25,48)$ tokens.
\end{example}


We see from the previous example that impermanent loss is not necessarily a ``loss'' in the literal sense. It refers to the situation where the present value of the deposit is greater than that of the pool. Thus, impermanent loss is often defined as the opportunity cost of providing liquidity (as opposed to simply holding the tokens). It is called ``impermanent'' because such a setback can be recovered once relative token price returns to its initial rate.

\subsection{Impermanent Gain}

Under CPMM, impermanent loss is always detrimental. This means, without a transaction fee, users are always better off holding onto their tokens. Alternatively, impermanent loss can be seen as the cost of maintaining a portfolio, where liquidity providers will always be selling the more valuable assets just to keep the weight of tokens balanced. It is the arbitragers who nip away the surplus value and leave behind a transaction fee. This, however, is not necessarily the case under the proposed protocol. Recall that a DEMM pool can be seen as an amalgamation of liquidity providers' portfolio preferences. In the event that a user's portfolio requires \emph{less} rebalancing maintenance than that of the pool, meaning that proportionately less value is taken away by arbitragers, such a profile could result in a relative gain. This unique phenomenon is named ``impermanent gain'' and will be explored in the next example.

\begin{figure}
\begin{center}
\begin{picture}(410,340)
\put(17,330){token $s$ = \$2 and token $t$ = \$10}
\put(250,330){token $s$ = \$1 and token $t$ = \$10}
\put(-5,265){Pool}
\put(33,265){\begin{tabular}{|c|c|}
\hline
40 token $s$ & 16 token $t$ \\
{[}\$80{]} & {[}\$160{]} \\
\hline
\end{tabular}}
\put(240,290){\begin{tabular}{|c|c|}
\hline
40 token $s$ & 16 token $t$ \\
{[}\$40{]} & {[}\$160{]} \\
\hline
\end{tabular}}
\put(240,240){\begin{tabular}{|c|c|}
\hline
$\approx63.50$ token $s$ & $\approx12.70$ token $t$ \\
{[}$\approx\$63.50${]} & {[}$\approx\$127.00${]} \\
\hline
\end{tabular}}
\put(164,267){\vector(3,1){75}}
\put(164,267){\vector(3,-1){75}}
\put(187,286){hold}
\put(178,237){DEMM}
\put(-5,155){Alice}
\put(39,155){\begin{tabular}{|c|c|}
\hline
20 token $s$ & 4 token $t$ \\
{[}\$40{]} & {[}\$40{]} \\
\hline
\end{tabular}}
\put(240,180){\begin{tabular}{|c|c|}
\hline
20 token $s$ & 4 token $t$ \\
{[}\$20{]} & {[}\$40{]} \\
\hline
\end{tabular}}
\put(240,130){\begin{tabular}{|c|c|}
\hline
$\approx31.75$ token $s$ & $\approx3.17$ token $t$ \\
{[}$\approx\$31.75${]} & {[}$\approx\$31.75${]} \\
\hline
\end{tabular}}
\put(164,157){\vector(3,1){75}}
\put(164,157){\vector(3,-1){75}}
\put(187,176){hold}
\put(178,127){DEMM}
\put(-5,45){Bob}
\put(33,45){\begin{tabular}{|c|c|}
\hline
20 token $s$ & 12 token $t$ \\
{[}\$40{]} & {[}\$120{]} \\
\hline
\end{tabular}}
\put(240,70){\begin{tabular}{|c|c|}
\hline
20 token $s$ & 12 token $t$ \\
{[}\$20{]} & {[}\$120{]} \\
\hline
\end{tabular}}
\put(240,20){\begin{tabular}{|c|c|}
\hline
$\approx31.75$ token $s$ & $\approx9.52$ token $t$ \\
{[}$\approx\$31.75${]} & {[}$\approx\$95.24${]} \\
\hline
\end{tabular}}
\put(164,47){\vector(3,1){75}}
\put(164,47){\vector(3,-1){75}}
\put(187,66){hold}
\put(178,17){DEMM}
\end{picture}
\caption{\label{f:one-sided} By net value, the pool's impermanent loss from Example \ref{ex:one-sided1} is $\approx-\$9.5$, Bob's impermanent loss is $\approx-\$13.01$, and Alice's impermanent gain is $\approx\$3.5$.}
\end{center}
\end{figure}
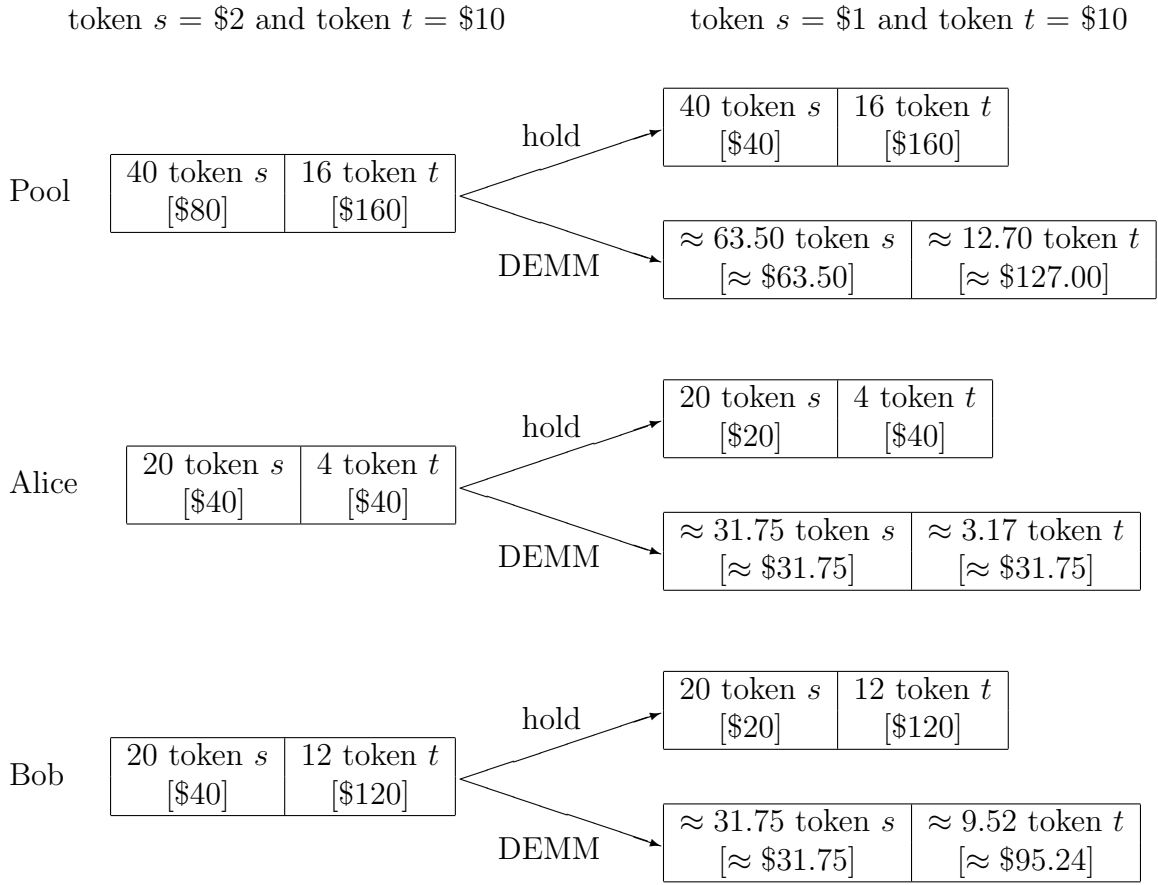

\begin{figure}
\begin{center}
\includegraphics{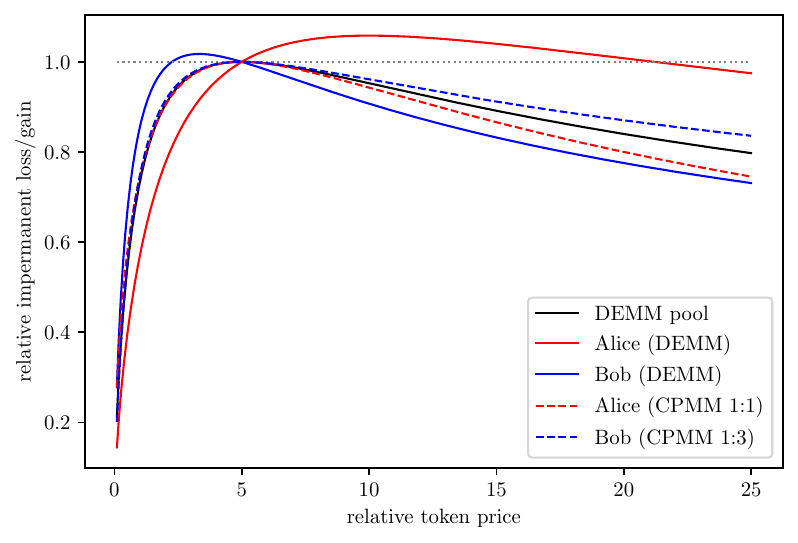}
\caption{\label{f:graph} Relative impermanent loss/gain from Example \ref{ex:one-sided1}}
\end{center}
\end{figure}

\begin{example}\label{ex:one-sided1}
We revisit Example \ref{ex:loss}, this time supposing that token prices are \$1 and \$10 respectively. Token values from the share of the pool and hold alternative are given in Figure \ref{f:one-sided}. Notice that Alice's allocation in the pool is greater in value than her hold alternative. This gain, however, does not come from nowhere. We can see that Bob's impermanent loss is greater than that of the pool by net value, and the difference is credited to Alice as a positive gain.

Next, Figure \ref{f:graph} plots relative impermanent loss/gain, i.e., $\frac{\textrm{total pool value}}{\textrm{total hold value}}$, against relative token price for the DEMM pool, Alice's share, Bob's share, together with their CPMM alternative. We would like to note that, unless the relative price is at its initial value, the pool itself is always at a relative loss under any AMM. Since everyone has tokens by the same weight in the predecessor protocol, their share of the pool can only be subject to depreciation through impermanent loss. While an overturn is possible for some individual investors under DEMM, it comes at the cost of a more severe impermanent loss in the case that token prices move in an unfavorable direction.
\end{example}

Impermanent gain can also happen when the tokens supplied by a partial liquidity provider depreciate against other tokens in the pool. In light of the earlier discussion, there is no ``surplus value'' to be taken away by arbitragers. Here, it is possible that the provider gains more of every deposited token.

\begin{example}\label{ex:one-sided2}
Consider again the pool $\chi'=((40,16),(1,2))$ from Example \ref{ex:rate}. Here, Carol holds 1 LP token $s$ together with 1 LP token $t$, and Dave holds 1 LP token $s$. Dave's initial investment is 8 token $s$, which is worth \$80.

Suppose now that the unit price of token $s$ has risen to \$64 while the unit price of token $t$ has decreased to \$5. Arbitragers will bring the pool to a new equilibrium state of $((2.5,64),(1,2))$. Note that Dave is now entitled to 32 token $t$ in this pool, which is more that what he has put in both in terms of token amount and net worth. This impermanent gain earned by Dave is caused by the fact that token $t$ depreciates against token $s$, resulting in an increased number of token $t$ in the pool.
\end{example}

Typically, impermanent loss becomes greater as relative token price moves away from its baseline. This holds for partial liquidity providers of DEMM as well. As we will see in the next example, they are exposed to token depletion in the case that the tokens they \emph{do not} supply lose value.

\begin{example}\label{ex:one-sided3}
Continuing from Example \ref{ex:one-sided1}, suppose that Carol redeems 0.2 LP token $s$ and 1 LP token $t$. The state of the pool would then be $\chi=((2,32),(0.8,1))$. Here, Carol and Dave own the entirety of token $s$ and $t$ in the pool respectively. Suppose that token $t$ loses value and its unit price becomes $\$\frac{5}{512}$ while token $s$ stands firm. This will change the state of the pool to $\chi=\left(\left(\frac{1}{16},512\right),(0.8,1)\right)$. Here, Carol's holding is reduced to $\frac{1}{16}$ token $s$.
\end{example}

While it may be interesting to develop the exact expression for impermanent loss (or gain) under DEMM, the formula will likely involve a great number of parameters and is beyond the scope of this paper.

\subsection{Token Addition and Removal}

It is possible for a dynamic exponent market maker protocol to operate with one inclusive pool, and there certainly are benefits and risks of doing so. From a trader's point of view, it is beneficial to be able to swap a token with any other token directly since that reduces transaction cost and gas fee. In addition, token exchange over a deeper pool results in a lower price impact, yielding a better exchange rate. On the other hand, as seen in the previous subsection, a pool with many assets exposes liquidity providers to a greater risk of impermanent loss.

Let $\chi=(\mathbf{r},\mathbf{w})$ be the state of the pool. We now outline two special operations that can be used to manipulate the inclusion of assets in the pool.

\begin{operation}[Add token]\label{o:add}
Let $t\in\{1,2,\ldots,n\}$. To add token $n+1$ to the pool, deposit $\Delta r_t$ token $t$ and $r_{n+1}$ token $n+1$ of equal value and update the state of the pool to $\chi'=(\mathbf{r'},\mathbf{w'})$ where
\[\mathbf{r'}=(r_1,r_2,\ldots,r_t+\Delta r_t,\ldots,r_n,r_{n+1})\]
and
\[\mathbf{w'}=\left(w_1,w_2,\ldots,\left(\frac{r_t+\Delta r_t}{r_t}\right)w_t,\ldots,w_n,\left(\frac{\Delta r_t}{r_t}\right)w_t\right).\]
Note that the depositor receives the same number of LP token $t$ and LP token $n+1$.
\end{operation}

\begin{operation}[Split pool]
Let $A\cup B$ be a partition of $\{1,2,\ldots,n\}$. An $A\cup B$ split results in two pools with state $\chi_A=((r_{t,t\in A}),(w_{t,t\in A}))$ and $\chi_B=((r_{t,t\in B}),(w_{t,t\in B}))$.
\end{operation}

These two operations should require consensus from the governance to perform. For instance, as we have seen in Example \ref{ex:one-sided3}, a malicious actor could add a depreciating token and drain values from the entire pool. On the other hand, pool splitting is a drastic and irreversible measure that disables trading between the two pools. Here, it does not affect existing liquidity providers since they may redeem their LP tokens from the respective pool. In this light, splitting away one asset is equivalent to removing that asset from the pool. Note also that if a token is completely withdrawn from the pool by liquidity providers, it cannot be resupplied using Operation \ref{o:provide2} (provide liquidity), and one must invoke Operation \ref{o:add} (add token) to reintroduce it.

\subsection{Flash Loan Attack}

In this subsection, we discuss a serious vulnerability of DEMM along with some possible safeguard mechanism. Since DEMM allows users to provide liquidity partially, an attacker can drain out a token for the purpose of providing liquidity in that particular token and securing an unfairly large share of that token in the pool. The process can be outlined as follows.
\begin{enumerate}
\item Swap a large amount token $s$ for token $t$.
\item One-sided deposit a small amount of token $t$.
\item Swap the remaining token $t$ for token $s$.
\item Redeem LP token $t$.
\end{enumerate}
Even without dynamic exponent, the attack described could threaten an automated market maker protocol that allows partial liquidity provision. What makes it especially catastrophic under DEMM is that partial liquidity provision manipulates the invariant function in a way that it favors the attack. We illustrate this in the next example.

\begin{example}\label{ex:attack}
Consider a DEMM pool with two tokens: token $s$ and token $t$. Suppose that the unit price of tokens $s$ and $t$ is \$5 and \$2 respectively, and the current state of the pool is $\chi=((4,10),(1,1))$. Initially, Eve has 36 token $s$ in her possession, and she performs the following transactions in order.
\begin{flushright}
\hfill resulting state of the pool
\end{flushright}
\begin{enumerate}
\item Swap 36 token $s$ for 9 token $t$. \hfill \makebox[\widthof{$\chi=((1.6,10),(1,2))$}][l]{$\chi=((40,1),(1,1))$}
\item Contribute 1 token $t$ into the pool and earn 1 LP token $t$. \hfill \makebox[\widthof{$\chi=((1.6,10),(1,2))$}][l]{$\chi=((40,2),(1,2))$}
\item Swap 8 token $t$ for 38.4 token $s$. \hfill $\chi=((1.6,10),(1,2))$
\item Redeem 1 LP token $t$ for 5 token $t$. \hfill \makebox[\widthof{$\chi=((1.6,10),(1,2))$}][l]{$\chi=((1.6,5),(1,1))$}
\end{enumerate}
At the end of the day, Eve walks away with a total of 38.4 token $s$ and 5 token $t$, reaping a net profit of 2.4 token $s$ and 5 token $t$.

Note that, had the invariant function stayed fixed at $f(x_s,x_t)=x_sx_t$ for the entire situation, Eve would have gotten only 32 token $s$ in the third step. In this case, Eve will end up with $(32,5)$ tokens, and ultimately it would seem as if she simply exchanges 4 token $s$ for 5 token $t$.
\end{example}


Another factor that makes this attack readily available is the existence of flash loan. Here, flash loan is an uncollateralized borrowing unique to decentralized finance where users borrow and return assets to the lender within the same blockchain transaction. Since lending out to a flash loan is risk-free, borrowers may temporarily get a hold of a large amount of funds. Hence, for the attack described, it is possible for an attacker to obtain a large amount of any particular token to begin with.

The essence of this vulnerability is that an attacker needs to provide liquidity while the pool, and hence the invariant function, is unnaturally skewed. The entire operation must be instantaneous, or else an arbitrager could correct the pool before the attacker could act. Therefore, one possible countermeasure is to delay state update during liquidity provision. Here, when a liquidity provider contributes to the pool, the fund is on hold for a random number of blocks before it officially enters the pool. As discussed in Example \ref{ex:attack}, this would discourage a straightforward flash loan attack since the invariant function will no longer shift in favor of the attack. Individuals with a large amount of capital may still attempt to manipulate the pool, but at the risk of losing them to the miners who have caught wind of the attack.

Alternatively, one may take an inspiration from Uniswap price oracle \cite{AZR,AZSKR} and use geometric time-weighted average for the numbers of LP tokens to be minted during liquidity provision. Let $\left(\mathbf{r}^{(-k)},\mathbf{w}^{(-k)}\right),\ldots,\left(\mathbf{r}^{(-1)},\mathbf{w}^{(-1)}\right),\left(\mathbf{r}^{(0)},\mathbf{w}^{(0)}\right)$ be the state of the pool at the beginning of the past $k$ blocks and the current one. For token $t$, $t\in\{1,2,\ldots,n\}$, we see that the geometric average of the number of LP token per actual token in the pool is
\[G_t=\left(\prod_{j=-k}^0\frac{w_t^{(-j)}}{r_t^{(-j)}}\right)^\frac{1}{k+1}.\]
It will be more difficult to manipulate the minting of LP tokens if this average ratio is used  for exponent update during liquidity provision. Namely, \eqref{e:provide2} from Operation \ref{o:provide2} can be given instead as
\[\mathbf{w'}=((r_1+\Delta r_1)G_1,(r_2+\Delta r_2)G_2,\ldots,(r_n+\Delta r_n)G_n).\]
Finally, we would like to remark that the effectiveness of the strategies considered remains a topic for future research, and the proposed protocol may still be subject to other unexplored vulnerabilities.


\section{Conclusion}

In this paper, we propose an automated market maker protocol that uses constant product formula with adaptive exponent. The invariant function becomes flexible during liquidity provision and welcomes deposits of assets in any proportion, including one-sided, and the pool maintains total token value ratio for each user through arbitraging. Here, a single pool can encompass all tradable assets and house all trading pairs, at the risk of impermanent loss. In conjunction with one-sided liquidity providing, an unheard-of impermanent gain could also occur. It is also possible to include new assets and remove assets from the pool on the fly. Finally, we discuss a known protocol vulnerability as well as some possible preventions.

\section*{Acknowledgments}
The author wishes to thank Nipun Pitimanaaree for several discussions and comments, and the anonymous reviewers for their insightful suggestions that have helped improve the paper.

\end{document}